\documentclass[aps,prl,twocolumn,superscriptaddress]{revtex4-1}

\usepackage{hyperref}
\usepackage{color}
\usepackage{graphicx,psfrag, subfigure}
\usepackage{color,soul}
\usepackage{tensor}

\usepackage{mathtools}
\usepackage{braket}

\usepackage{array}
\usepackage{amssymb}
\usepackage{amsmath}
\DeclareSymbolFont{bbold}{U}{bbold}{m}{n}
\DeclareSymbolFontAlphabet{\mathbbold}{bbold}
\usepackage{color}
\usepackage{eufrak}
\usepackage{graphicx,psfrag, subfigure}
\usepackage{hyperref}
\DeclareMathAlphabet{\mathpzc}{OT1}{pzc}{m}{it}
\usepackage{blkarray}

\usepackage{yfonts}
\usepackage{mathrsfs}
\usepackage{relsize}
\usepackage{tabularx}

\DeclareMathAlphabet\mathbfcal{OMS}{cmsy}{b}{n}

\def\be{\begin{equation}}
\def\ee{\end{equation}}
\def\bea{\begin{eqnarray}}
\def\eea{\end{eqnarray}}

\def\be{\begin{equation}}
\def\ee{\end{equation}}
\def\bea{\begin{eqnarray}}
\def\eea{\end{eqnarray}}
\usepackage{dsfont}
\usepackage[mathscr]{eucal}

\newcommand{\bs}{\boldsymbol}

\def\simleq{\; \raise0.3ex\hbox{$<$\kern-0.75em
      \raise-1.1ex\hbox{$\sim$}}\; }

\def\simgeq{\; \raise0.3ex\hbox{$>$\kern-0.75em
      \raise-1.1ex\hbox{$\sim$}}\; }

\begin{document}

\title{Testing the electric Aharonov-Bohm effect with superconductors}

\author{Thomas C. Bachlechner}
\affiliation{Department of Physics, University of California San Diego, La Jolla, USA}
\author{Matthew Kleban}
\affiliation{Department of Physics, New York University, New York, USA}
\vskip 4pt

\begin{abstract}

{The phase of the wave function of charged matter is sensitive to the value of the electric potential, even when the matter never enters any region with non-vanishing electromagnetic fields. Despite its fundamental character, this archetypal electric Aharonov-Bohm effect has evidently never been observed. We propose an experiment to detect the electric potential through its coupling to the superconducting order parameter.  A potential difference between two superconductors  will induce a relative phase shift that is observable via the DC Josephson effect even when no electromagnetic fields ever act on the superconductors, and even if the potential difference is later reduced to zero.  This is a type of electromagnetic memory effect, and would directly demonstrate the physical significance of the electric potential.}

\end{abstract}

\maketitle

{\it Introduction}~---~
 Electrodynamics is conventionally described using scalar and vector potentials, even though in classical physics only the electric and magnetic field strengths are observable.  It was noticed long ago that the potentials themselves have direct physical significance in that they can affect the phase of the quantum mechanical wave function of charged matter even in regions where the field strengths vanish \cite{Ehrenberg_1949,PhysRev.115.485,PhysRev.123.1511}.   Such phase differences can then be observed with interference experiments.

 In their seminal paper \cite{PhysRev.115.485}, Aharonov and Bohm describe two archetypal versions of their effect.  The best known version today is magnetostatic, with vanishing electric field and a magnetic field $\bf B$ that is nonzero only within a solenoidal tube.  In this situation the  vector potential is (necessarily) non-vanishing \footnote{More precisely, there must be a  component with non-zero curl.} outside the tube, and charged particles that propagate around it will experience a phase shift proportional to the magnetic flux in the tube, despite never entering the region of nonzero $\bf B$.  This magnetic Aharonov-Bohm (AB) effect  was observed long ago \cite{1960PhRvL...5....3C,1962NW.....49...81M}.  
 
 The obvious electric counterpart is a setup where charged particles propagate only in regions of vanishing electric field $\bf E$, but different potential due to the presence of nonzero electric fields somewhere else.  The experiment proposed in \cite{PhysRev.115.485} was to pass two electron beams through Faraday cages, with a different time-varying voltage applied to each.  To our knowledge this experiment has never been carried out, nor has the electric version of the AB effect ever been experimentally verified in any other way.  
 
 In this work we propose a simple and feasible experimental setup employing superconductors.  This could verify for the first time the physical significance of the electric potential in a region of vanishing fields. For our purposes it is essential that the charged particles are never exposed to non-zero electric fields. Such an effect -- where the interaction is between charges and potentials in a region of vanishing field strengths -- is sometimes referred to as Aharanov-Bohm  type I, while effects that can be explained by interactions in regions of non-vanishing field strengths are referred to as type II \cite{1996fqml.conf....8A}. There have been some experimental studies of the electric AB effect that failed to verify the effect \cite{PhysRevB.40.3491}, while others made positive observations  \cite{1998Natur.391..768V,PhysRevB.67.033307}.  However, these two experiments constitute measurements of the Type II effect in that the charged particles traversed regions of non-zero electric field.  In this work we are interested exclusively in the type I effect that (to our knowledge) has never been tested \footnote{Several references \cite{PhysRevLett.59.1791, 1998Natur.391..768V} refer to an article in a conference proceedings \cite{mysteryreference}.  Unfortunately we have not been able to retrieve this article, and no details are given in any of the papers that cite it.}.
 

\begin{figure}[t]
\centering
\includegraphics[width=.48\textwidth]{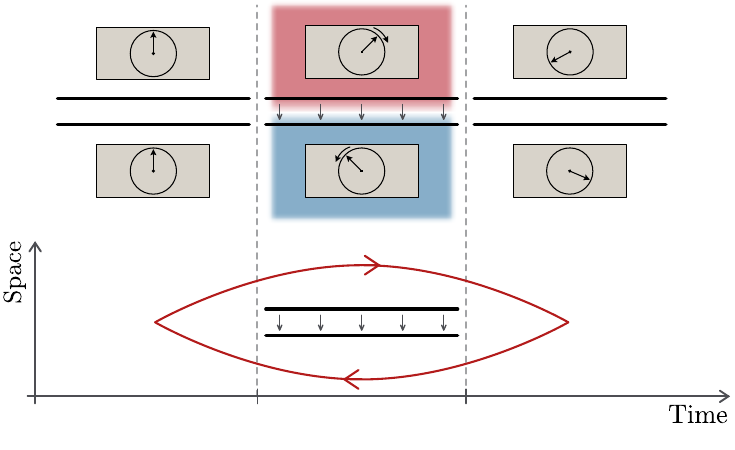}
\caption{\small 
Basic setup (top) and contour of Wilson loop (bottom). An electric potential difference $V$ maintained for a time $T$ across two superconductors induces a relative superconducting phase  difference $\Delta \theta=q  VT/\hbar$.}\label{basic}
\end{figure}

\bigskip

Consider two  superconductors that are initially connected so that their phase difference is zero, $\Delta \theta = 0$.  Subsequently, the superconductors are placed on either side of a large planar capacitor.  In the temporal gauge (where $\bs E=-\dot{\bs A}$) Gauss' law reads
\be
\rho=-\bs\nabla \cdot \dot{\bs A}\,.
\ee
Charging the capacitor and maintaining a fixed voltage across it changes the initially vanishing vector potential, $\bs A (t=0) =\bs 0$, to a pure gradient, $\bs A(t)=\bs \nabla\lambda(x,t)$, that is non-vanishing between the capacitor plates. Assuming the capacitor plates are very large or that the superconductors are enclosed in Faraday cages, the electric field at the superconductors remains zero at all times.  The function $\lambda(x,t)$  depends linearly on time while a constant voltage on the capacitor is maintained, and is independent of position outside but sufficiently near the capacitor plates and inside any Faraday cages.\footnote{In the approximation that the capacitor plates are infinitely large and located at $x=0$ and $x=L$, one simply has
$\lambda(t,x) = 0$ for $x<0$,  $\lambda(t,x) = (x/L) \int V(t) dt$ for $0 < x < L$, and $\lambda(t,x) = \int V(t) dt$ for $x>L$.}

We can eliminate the non-trivial vector potential by the gauge transformation $\bs A\rightarrow \bs A-\bs \nabla\lambda(x,t)$, where again $\lambda(x,t)=\int dx  A_{x}$ is constant in space but takes different values on either side of the capacitor. The gauge transformation acts on the phase of the superconductors as $\theta\rightarrow \theta-q \lambda/\hbar$, showing that a time-dependent gauge configuration $\lambda \propto t$ is equivalent to $\Delta \theta \propto t$ -- a phase difference that increases in proportion to  time and to the voltage on the capacitor.   This  is illustrated in Figure \ref{basic}. 

Once the capacitor is discharged the time-dependence disappears and the phase difference becomes constant.  As long as the superconductors remain isolated from any further influences, this phase difference remains eternally imprinted, and can (in principle) be observed at any later time by re-connecting the superconductors and using the DC Josephson effect.  
This is an example of  ``electromagnetic memory" \cite{Bieri:2013hqa,Susskind:2015hpa,Pasterski2017}.  Together with  more general question of the physical significance and infrared dynamics of gauge potentials, it  is relevant for many questions of interest to ongoing research,  including soft theorems (see \cite{Strominger:2017zoo} for a recent pedagogical review) and is related to gravitational memory and black hole information loss \cite{Hawking:2016msc}.

Taken together, the trajectories of the superconductors trace out a closed loop in space-time that we illustrate in Figure \ref{basic} -- initially connected, then separated to either side of the capacitor, and then again connected.  The Wilson loop integral $\oint A_\mu dx^\mu$ (with $A_\mu$ the four-potential) is gauge invariant and non-zero when taken along this loop. 

{\it Previous work}~---~The  experiment originally proposed by Aharonov and Bohm to measure the electric version of their eponymous effect was an electron interference experiment where a time-dependent   voltage $V$ is applied to the exterior of two Faraday cages  while the electron beams pass through them \cite{PhysRev.115.485}.   If the voltage difference is non-zero only during the time $T$ when the electrons are well contained inside the cages, the electrons never enter a region with non-zero electric field.  Nevertheless, applying a different potential to each cage would induce a relative phase shift of 
\be
\Delta \theta= {q \Delta V T\over \hbar}\,,
\ee 
in the electron wave-functions, where $V$ coincides with $A_0$ in Coulomb gauge.  The beams can then be interfered in order to measure the phase shift.  However  in order to prevent the electrons from entering a region of non-zero fieds, the typically high velocity of electron beams would require the voltage  to be switched on and off extremely rapidly, over times of order $ 10^{-9}\text{s}$.  This is challenging  and would induce strong non-adiabatic fields \cite{Badurek, Schutz:2013fca}. An interesting application of the electric Aharonov-Bohm effect was proposed in \cite{Arvanitaki:2007gj}, but so far no positive observations have been made.   

Instead of using charged particles directly, in this work we  propose to employ the superconducting Cooper pair condensate in a quantum interference experiment that is sensitive to the electric potential in a region of vanishing fields. Using superconductors allows a more experimentally feasible test of the significance of the scalar potential.  A related gedanken-experiment was proposed in \cite{Susskind:2015hpa}.

Reference \cite{PhysRevLett.108.230404} proposed an experiment measure the relative phase shift between two superconductors induced a  gravitational potential difference, analogous to the electric Aharonov-Bohm effect we discuss here.

\begin{figure}[t]
\centering
\includegraphics[width=.38\textwidth]{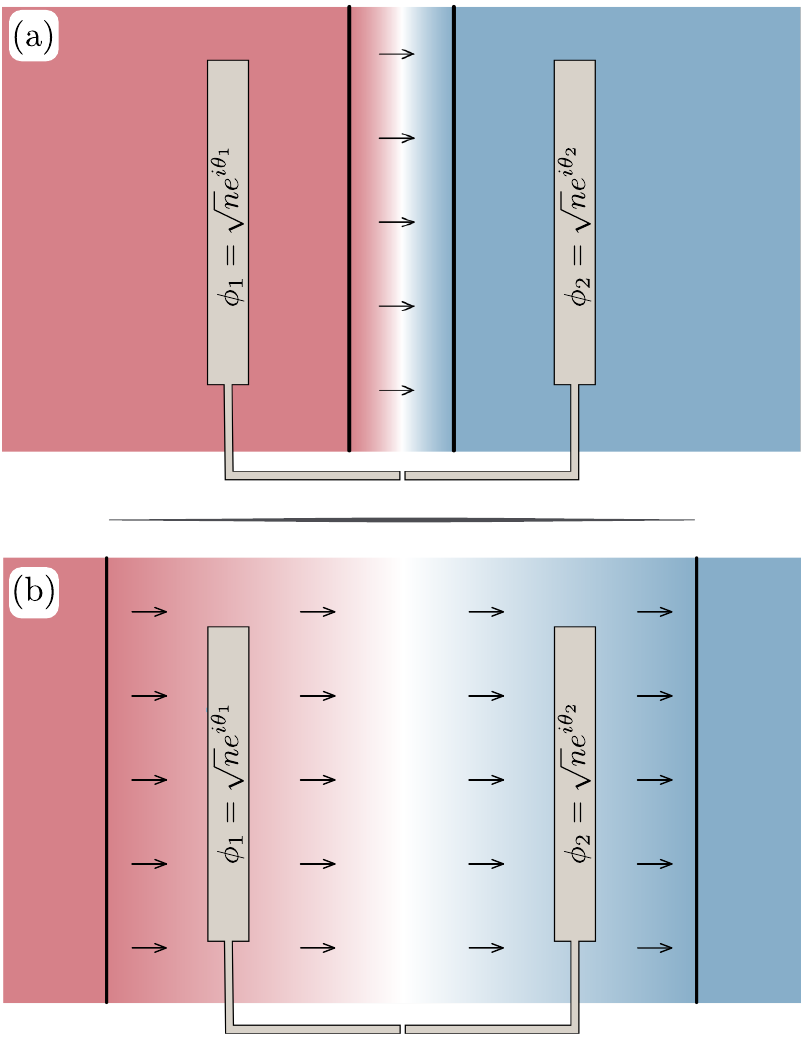}
\caption{\small Two superconductors subject to an electric potential difference $V$ and connected by a thin superconducting wire with a  junction.  Positive/negative electric potential is illustrated as red/blue shading, and the arrows indicate the magnitude and direction of the field. {\it (a)} A charged capacitor is in between the two superconductors, creating a potential difference but zero electric field on the superconductors. {\it (b)} The capacitor encloses the two superconductors, creating a field that acts directly on the superconductors. }\label{exp}
\end{figure}

\bigskip
{\it Experimental setup}~---~The experiment consists of two superconductors that are each embedded between large planar capacitors as shown in Figure \ref{exp}. The superconductors are described by the Ginzburg-Landau order parameters $\phi_{1,2}=\sqrt{n}e^{i\theta_{1,2}}$, where $n$ denotes the constant Cooper pair density and $\theta_{1,2}$ are the time-dependent phases. Thin superconducting wires originate from each of the superconductors and terminate at an insulating junction located far  from the capacitors. The  junction contains a phase-dependent gradient of the order parameter that allows Cooper pairs to tunnel. This leads to an observable supercurrent described by the first Josephson relation 
$$I=I_\text{c}\sin(\Delta \theta),$$
where $I_\text{c}$ is the critical current that depends on the detailed configuration of the junction \cite{JOSEPHSON1962251,RevModPhys.36.216,1974RvMP...46..251J}. 

We present two distinct experimental setups corresponding to different configurations of  the capacitor plates, labeled by (a) and (b) in Figure \ref{exp}. In configuration (a) a voltage difference  is induced between the two superconductors, which lie in regions of vanishing field.  In configuration (b) there is both a non-zero voltage difference and a non-vanishing electric field acting on the superconductors.

Both setups result in an electric potential difference of $V(t)$ between the superconductors. The Cooper pairs carry a charge $q=-2e$, so the gauge-coupling term in the action, $S\supset{2e\over \hbar}\oint A_\mu dx^\mu$ (with $A_\mu$ the four-potential), yields an associated phase shift given by the second Josephson relation, 
$$\Delta \theta={2e\over \hbar}\int V(t)\,dt. $$ 
The path $V(t)$ is imprinted as a memory in the relative phase.
The electric field at the superconductors vanishes in setup (a), but acts locally in setup (b). Therefore, experimental verification of the second Josephson relation in setup (a) acts as an indication that the scalar potential is physical (type I electric Aharanov-Bohm), while setup (b)  represents a more conventional Josephson setup junction and could serve as a control  (type II electric Aharanov-Bohm).
In both setups the relation between the relative phase and the voltage difference is identical, but the  electric field configurations differ. 



A constant voltage of $1\mu V$ will shift the frequency of an AC Josephson current by about $5\text{GHz}$. An application of a short voltage pulse of $10\text{nV}$ over a period of $10\text{ns}$ would shift the relative phase by $\pi$.   A precise measurement this phase-shift may be difficult due to the  small timescales and voltages involved, but any phase shift induced by the configuration in  Figure \ref{exp} (a) would demonstrate the existence of the type I electric Aharanov-Bohm effect. 


There are several systematic effects that can impact the observations or their interpretations. First, in setup (a) we attempted to place the Cooper pair condensate in a region that is entirely  separated spatially from the region of non-vanishing electric fields. However, since we also need to observe a current and moving the superconductors is difficult, we rely on superconducting wires leading to the junction. These wires will experience the non-vanishing fringe fields of the capacitor. Phase-coherence is maintained within each superconductor, and the number density of Cooper pairs is spatially constant, so we can separate the Ginzburg-Landau action for the condensate into two additive contributions: one from the wires and one from the bulk of the superconductors. The phase couples linearly to the potentials, so if we denote by $\epsilon={\cal V}_\text{wire}/{\cal V}_\text{bulk}$ the relative volume between the wire and the bulk of the superconductor, we expect the systematic error from fringe fields in the second Josephson relation to be no larger than $\epsilon$. Assuming superconductors of volume $1\text{cm}^3$, the volume ratio can be as small as $\epsilon \gtrsim 10^{-11}$ with $\mu \text{m}$-scale wires. Conversely, if electric fields instead of  potentials were fundamental, the second Josephson relation would lead to a phase velocity that is multiplied by the small factor of $\mathcal{O}(\epsilon)$. If realized in nature, a positive direct and unambiguous observation of this significantly weaker effect would actually be easier. 

Second, thermal noise will induce non-vanishing electric fields within the Faraday cages. Assuming the conductivity of copper and a frequency bandwidth of $10^6 \text{Hz}$, a $1\text{cm}$ Faraday cage contains thermal voltage fluctuations of around $10^{-11}\text{V}$. These fluctuations induce a negligible relative voltage difference between the superconductors.

{\it Conclusions}~---~A positive observation would provide the first experimental evidence for the electric Aharanov-Bohm effect. Conversely, a  negative observation ruling this effect out would be of profound importance for our understanding of quantum gauge theories and consistent with a holonomic theory of quantum electrodynamics \cite{tbtoappear}. Either observation would be of elementary importance.

{\it Acknowledgements}~---~
We thank Mina Arvanitaki, Robert Dynes, Dan Green, Raphael Flauger, Per Kraus, Liam McAllister, John McGreevy,  Javad Shabani, and Ken Van Tilburg for useful discussions. The work of TB was supported in part by DOE under grants no. DE-SC0009919 and by the Simons Foundation SFARI 560536. The work of MK is supported by the NSF through grants PHY-1214302 and PHY1820814.  This work was performed in part at the Aspen Center for Physics, which is supported by National Science Foundation grant PHY-1607611.

\bibliographystyle{apsrev4-1}

\bibliography{bubblerefs.bib}

\end{document}